\documentclass[reqno]{amsart}
\usepackage{amsmath,amsfonts,amssymb}
\usepackage{verbatim}
\usepackage{enumerate}
\usepackage{url}
\def\N{\mathbb N}

\def\Z{\mathbb Z}

\def\F{\mathbb F}
\def\ord{\mathop{\rm ord}\nolimits}
\def\rad{\mathop{\rm rad}}

\theoremstyle{plain}
\newtheorem{theorem}{Theorem}[section]
\newtheorem{lemma}[theorem]{Lemma}
\newtheorem{definition}[theorem]{Definition}
\newtheorem{corollary}[theorem]{Corollary}

\newtheorem{remark}[theorem]{Remark}
\newtheorem{example}[theorem]{Example}
\def\proof{{\it Proof: }}
\def\qed{\hfill\hbox{$\square$}}

\theoremstyle{definition}

\numberwithin{equation}{section}

\author[F.E. Brochero Mart\'{\i}nez]{F. E. Brochero Mart\'{\i}nez}
\author[C. R. Giraldo Vergara]{C. R. Giraldo Vergara}
\address{
Departamento de Matem\'{a}tica\\
Universidade Federal de Minas Gerais\\
UFMG\\
Belo Horizonte, MG\\
 30123-970\\
 Brazil\\
 }
 \email{fbrocher@mat.ufmg.br }\email{carmita@mat.ufmg.br}

\date{\today
}

\subjclass[2000]{}

\subjclass[2010]{ 12E05(primary) and 94B05(secondary)} 
\title{Weight enumerator of some irreducible cyclic codes}

\keywords{Cyclic Codes, Weight Enumerator, minimal distance}
\begin{document}
\maketitle

\begin{abstract} 
In this article, we show explicitly all possible weight enumerators for every irreducible cyclic code of length $n$ over a finite field $\F_q$,  in the case which each prime divisor of $n$ is also a divisor of $q-1$.  

\end{abstract}

\section{Introduction}

A   code of lenght $n$ and dimension $k$  over a finite field $\F_q$ is a linear $k$-dimensional subspace of $\F_q^n$. 
A $[q;n,k]$-code  $\mathcal C$ is called {\em cyclic}  if  it is  invariant by the shift permutation, i.e.,  if $(a_1,a_2,\dots,a_n)\in \mathcal C$ then the shift $(a_n,a_1,\dots,a_{n-1})$ is also in $\mathcal C$. 
The cyclic code $\mathcal C$ can be viewed as an ideal in the group algebra $\F_qC_n$,
where $C_n$ is the cyclic group of order $n$. We note that $\F_qC_n$ is isomorphic  to $\mathcal R_n=\frac{\F_q[x]}{\langle x^n-1\rangle}$
 and since subspaces of $\mathcal R_n$  are ideals and   $\mathcal R_n$ is a principal ideal domain,  it follows that  each ideal  is generated by a polynomial $g(x)\in \mathcal R_n$, where  $g$ is a divisor of $x^n-1$.  

Codes generated by a polynomial of the form $\frac{x^n-1}{g(x)}$, where $g$ is an irreducible factor of $x^n-1$, are called  {\em minimal cyclic codes}. Thus, each minimal cyclic code is associated of  natural form with an irreducible factor of $x^n-1$ in $\F_q[x]$.  
 An example of minimal cyclic code is  the  Golay code that was used on the Mariner Jupiter-Saturn Mission (see \cite{Gol}),  the BCH code used in  communication systems like VOIP telephones and
 Reed-Solomon code used in two-dimensional bar codes and  storage  systems  like compact disc players, DVDs, disk drives, etc (see  \cite[Section 5.8 and 5.9]{FaCa}).  The advantage of the cyclic codes, with respect to other linear codes,  is that they have efficient encoding and decoding algorithms (see  \cite[Section 3.7]{FaCa}).

For each element of $g\in \mathcal R_n$, $\omega(g)$ is defined as the number of non-zero coefficient of $g$ and is called the {\em Hamming  weight of the word $g$}. 
 Denote by  $A_j$ the number of codewords with weight $i$   and by $d=\min\{i>0| A_i\ne 0\}$  the minimal distance of the code.  A $[q;n,k]$-code with minimal distance $d$ will be denoted by $[q;n,k,d]$-code. 
The sequence $\{A_i\}_{i=0}^n$ is called the {\em weight distribution} of the code and $A(z):=\sum_{i=0}^n A_iz^i$ is  its {\em weight enumerator}.  The importance of the weight distribution  is that it allows us to measure the probability of non-detecting an error of the code: For  instance, the probability of  undetecting an error in a binary symmetric channel is
$\sum\limits_{i=0}^n A_i p^i(1-p)^{n-i}$,
where $p$ is the probability  that, when the transmitter  sends  a binary symbol ($0$ or $1$),  the  receptor gets the wrong symbol.

The weight distribution of  irreducible cyclic codes has been determined for a small number of special cases. For a survey about this subject  see \cite{Din}, \cite{DiJi} and their references.

In this article, we show all the possible  weight distributions of length $n$ over a finite field $\F_q$ in the case which  every prime divisor of $n$ divides $q-1$.

\section{Preliminaries }
Throughout this article, $\F_q$ denotes a finite field of order $q$, where $q$ is a power of a 
 prime,  
$n$ is a positive integer such that
$\mbox{gcd}(n,q)=1$,
$\theta$  is a generator of the cyclic group $\F_q^*$ and $\alpha$ is a generator of the cyclic group $\F_{q^2}^*$ such that $\alpha^{q+1}=\theta$. 
For each $a\in \F_q^*$, $\ord_q a$  denotes  the minimal positive integer $k$ such that $a^k=1$, for each prime  $p$   and each integer $m$, $\nu_p(m)$  denotes the maximal power of $p$ that divides $m$ and  $\rad(m)$ denotes the radical of $m$, i.e.,  
 if $m=p_1^{\alpha_1}p_2^{\alpha_2}\cdots p_l^{\alpha_l}$ is the factorization of $m$ in prime factors, then $\rad(m)=p_1p_2\cdots p_l$.  
Finally,   $a_{_{\div b}}$ denotes the integer $\frac a{\gcd(a,b)}$.

Since each irreducible factor of $x^n-1\in \F_q[x]$ generates an irreducible cyclic code of length $n$,
then a fundamental problem of code theory is to characterize these irreducible factors.
The problem of  finding  a ``generic algorithm''  to split   $x^n-1$ in $\F_q[x]$, for any $n$ and $q$, is  an open one and only some particular cases are known. 
Since $x^n-1=\prod_{d|n} \Phi_d(x)$, where $\Phi_d(x)$ denotes  the $d$-th cyclotomic polynomial (see \cite{LiNi} theorem 2.45), it follows that the factorization of $x^n-1$ strongly  depends on the factorization of the cyclotomic polynomial that  has been studied by several authors (see \cite{Mey},  \cite{FiYu}, \cite{WaWa} and \cite{CLT}).   

In particular, a natural question is to find conditions in order to  have  all the  irreducible factors    binomials or trinomials. 
In this direction,  some results are  the following ones

\begin{lemma}\cite[Corollary 3.2]{BGO} \label{factorsb}
Suppose that
\begin{enumerate}
\item $\rad(n)|(q-1)$ and
\item $8\nmid n$ or $q\not\equiv 3 \pmod 4$.
\end{enumerate}
Then the factorization of $x^n-1$ in irreducible factors of $\F_q[x]$ is
$$\prod_{t|m}\prod_{{1\le u\le \gcd(n,q-1)\atop \gcd(u,t)=1}} (x^t-\theta^{ul}),$$
where $m=n_{_{\div (q-1)}}$ and $l=(q-1)_{_{\div n}}$.
In addition,  for each  $t$ such that $t|m$, the number of irreducible factors of degree $t$ is $\frac{\varphi(t)}t\cdot \gcd(n,q-1)$,  where $\varphi$ denotes the Euler Totient function.
\end{lemma}

\begin{lemma}\cite[Corollary 3.4]{BGO} \label{factorst}
Suppose that
\begin{enumerate}
\item $\rad(n)|(q-1)$ and
\item $8\mid n$ and $q\equiv 3 \pmod 4$.
\end{enumerate}
Then the factorization of $x^n-1$ in irreducible factors of $\F_q[x]$ is
$$\prod_{t|m'\atop t\text{ odd}}\prod_{{1\le w\le\gcd(n,q-1)}\atop \gcd(w,t)=1} (x^t-\theta^{wl})\cdot \prod_{t|m'}\prod_{ u\in\mathcal S_t} (x^{2t}-(\alpha^{ul'}+\alpha^{qul'})x^t+\theta^{ul'}),$$
where $m'=n_{_{\div (q^2-1)}}$ and $l=(q-1)_{_{\div n}}$, $l'=(q^2-1)_{_{\div n}}$, 
$r=\min\{\nu_2(\frac n2),\nu_2(q+1)\}$ and   $\mathcal S_t$ is the set 
 $$\left\{u\in\N\left|  {1\le u\le\gcd(n, q^2-1), \gcd(u,t)
=1\atop 
2^r\nmid u \text{ and }\ u< \{qu\}_{\gcd(n,q^2-1))}}\right.\right\},$$
where $\{a\}_b$ denotes the  remainder of the division of $a$ by $b$, i.e.  it is  the number $0\le c<b$ such that $a\equiv c\pmod b$ .

In addition, for each $t$  odd such that $t|m'$, the number of irreducible binomials  of degree $t$ and   $2t$ is
$\dfrac{\varphi(t)}t\cdot \gcd(n,q-1)$ and
$\dfrac{\varphi(t)}{2t}\cdot \gcd(n,q-1)$    respectively,  and
the number irreducible trinomials  of degree $2t$ is
$$
\begin{cases}
\dfrac{\varphi(t)}t\cdot 2^{r-1} \gcd(n,q-1),&\text{if $t$ is even}\\
\dfrac{\varphi(t)}{t}\cdot (2^{r-1}-1) \gcd(n,q-1),&\text{if $t$ is odd}.
\end{cases}
$$
\end{lemma}

\section{weight distribution}
Throughout  this section, we  assume that $\rad (n)$ divides $q-1$ and $m$, $m'$ $l$, $l'$ and $r$ are as in the lemmas \ref{factorsb} and \ref{factorst}.
The following results characterize all the possible cyclic codes of  length $n$ over $\F_q$ and show explicitly the weight distribution in each case. 

\begin{theorem}\label{4k+1} 
  If $8\nmid n$ or $q\not\equiv 3 \pmod 4$, then every irreducible code of length $n$ over $\F_q$   is a $[q;n, t, \frac nt]$-code where $t$ divides  $m$ 
and its weight enumerator is 
$$A(z)=\sum_{j=0}^{t}  {t\choose j}(q-1)^j z^{j\frac nt}=(1+(q-1)z^{\frac nt})^t.$$
\end{theorem}

\proof As a consequence of  Lemma \ref{factorsb}, every irreducible factor of $x^n-1$ is of the form $x^t-a$ where $t|n$ and $a^{n/t}=1$, so every irreducible code $\mathcal C$ of lenght $n$ is generated by a polynomial of the form
$$g(x)=\frac{x^n-1}{x^t-a}=\sum_{j=0}^ {n/t-1} a^{\frac nt-1 -j} x^{t j}$$
and $\{g(x), xg(x), \dots, x^{\frac nt-1}g(x)\} $ is a base of the $\F_q$-linear subspace $\mathcal C$.  Thus, every codeword in $\mathcal C$ is of the form $a_0g+a_1xg+\cdots +a_{t -1} x^{t-1}g $, with $a_j\in \F_q$,   and
$$\omega(a_0g+a_1xg+\cdots +a_{t -1} x^{t-1}g)= \omega(a_0g)+ \omega(a_1xg)+\cdots + \omega(a_{t -1} x^{t-1}g).$$
Since $\omega(g)=\frac nt$, it follows that
$$\omega(a_0g+a_1xg+\cdots +a_{t -1} x^{t-1}g)=\frac nt \#\{j| a_j\ne 0\}.$$
Clearly we have
$  {A}_k =  0$ for all $k$ that is not divisible by $\frac nt$. On the other hand, if $k = j\frac n t$, then exactly $j$ elements of this  base have non-zero coefficients  in the linear combination and each non-zero coefficient  can be chosen of $ q-1 $ distinct forms, hence
 $ {A}_k = {t \choose j} (q-1) ^ j. $
Then the weight distribution is 
$$ A_k = \begin{cases} 0, & \hbox{if }  t \nmid k \\
              {t \choose j}(q-1)^j, & \hbox{if }k = j\frac nt
\end{cases}
$$
as we want to prove. \qed
\begin{remark} The previous result generalizes Theorem 3 in \cite{ShBa} (see also Theorem 22 in \cite{DiJi}).
\end{remark}

\begin{remark}
As a direct consequence of Lemma \ref{factorsb},  for all  $t$ positive divisor  of  $m$, 
there exist $\frac {\varphi(t)}{t} \gcd(n,q-1)$ irreducible cyclic $[q;n,t,\frac nt]$-codes. 
\end{remark}

In order to find  the weight distribution in the case which $q\equiv 3 \pmod 4$ and $8|n$, we need some additional lemmas.

\begin{lemma}\label{weight}

Let  $t$ be a positive integer such that $t$ divides $m'$ and assume that   $q\equiv 3 \pmod 4$. 
If  $x^{2t}-(a+a^q)x^t+a^{q+1} \in \F_q[x] $ is  an irreducible trinomial, where $a=\alpha^{ul'}\in \F_{q^2}$, and $g(x)$ is  the polynomial 
 $ \dfrac{x^{n}-1}{x^{2t}-(a + a^q)x^t + a^{q+1}}\in\F_q[x]$,
then $\nu_2(u)\le r-2$ and 
$$\omega(g(x)-\lambda x^tg(x))=\begin{cases} 
\frac nt\left(1-\frac1{2^{r-\nu_2(u)} }\right),&\text{if $\lambda\in \Lambda_u$}\\
\frac nt,&\text{if $\lambda\notin \Lambda_u$,}
\end{cases}
$$
where $\Lambda_u=\left\{\left.\dfrac{a^{i}-a^{qi}}{a^{i+1} -  a^{q(i+1)}} \right| i=0,1,\dots, 2^{r-\nu_2(u)}-2\right\}$.
\end{lemma}

\proof 
Since $x^{2t}-(a+a^q)x^t+a^{q+1} $ is an irreducible trinomial in $\F_q[x]$, then $\gcd (t,u)=1$,  
$2^r\nmid u$ and $a\ne -a^q$. In particular, $\ord_{q^2} a$ does not divide either $q-1$ or $2(q-1)$.
Observe that 
\begin{align*}\ord_{q^2} a&=\frac {q^2-1}{\gcd( q^2-1, ul')}\\
&=\frac {q^2-1}{\gcd\left( q^2-1, u \frac {q^2-1}{\gcd(q^2-1,n)}\right)}\\
&=\frac {\gcd(q^2-1, n)}{\gcd(q^2-1,n,u)}\\
&=\frac {2^r\gcd(q-1, n)}{\gcd( 2^r(q-1),n,u)},
\end{align*}
and for each odd prime $p$,  we have
\begin{equation}\nu_p\left(\frac {2^r\gcd(q-1, n)}{\gcd( 2^r(q-1),n,u)}\right)\le \nu_p(\gcd(q-1, n))\le \nu_p(q-1).\label{power_p}
\end{equation}
Therefore
\begin{equation}
\nu_2 \left(\frac {2^r\gcd(q-1, n)}{\gcd( 2^r(q-1),n,u)}\right)=r+1-\nu_2(u)> \nu_2(2(p-1))=2.\label{power_2}
\end{equation}
On the other hand
\begin{align*}
g(x) &= \frac{x^{n}-1}{x^{2t} - (a + a^q)x^t + a^{q+1}}\\
     &= \frac{x^{n}-1}{a-a^q}\left(\frac{1}{x^t-a} - \frac{1}{x^t-a^q} \right) \\
&= \displaystyle{\sum_{j=1}^{n/t-1}\left( \frac{a^{j}-a^{qj}}{a-a^q}\right) x^{ n-t-tj} },\\
\end{align*}
is a polynomial whose degree is $n-2t$ and every non-null monomial is such that its degree is divisible by $t$.
  Now,  suppose that there exist $1\le i<j\le \frac nt -2$ such that the coefficients of  the monomials $x^{n-t-jt}$ and $x^{n-t-it}$   in   the polynomial  $g_{\lambda}:=g(x)-\lambda x^t g(x)$ are simultaneously  zero. Then
$$
\frac{a^{j} - a^{qj}}{a - a^q} = \lambda\frac{a^{j+1} - a^{q(j+1)}}{a - a^q} \quad\hbox{and}\quad
\frac{a^{i} - a^{qi}}{a - a^q} = \lambda\frac{a^{i+1} - a^{q(i+1)}}{a - a^q} .
$$
So, in the case which $\lambda\ne 0$, we have
$$ \lambda = \frac{a^{j}-a^{qj}}{a^{j+1} -  a^{q(j+1)}} = \frac{a^{i}-a^{qi}}{a^{i+1} -  a^{q(i+1)}}.$$
This last equality is equivalent to $ a^{(q-1)(j-i)} = 1$, i.e., 
 $\ord_{q^2} a$ divides $(q-1)(j-i)$. 
In the case $\lambda=0$,  we obtain that $\ord_{q^2} a$ divides $(q-1)j$ and $(q-1)i$ by the same argument.
 Thereby, we can treat this case as a particular case of the above one making $i = 0$.
It follows that $\frac {2^r\gcd(q-1, n)}{\gcd( 2^r(q-1),n,u)}$ divides $(q-1)(j-i)$. 

So,  by the Equation (\ref{power_p}), the condition $\ord_{q^2} a|(q-1)(j-i)$ is equivalent to 
$$\nu_2 \left(\frac {2^r\gcd(q-1, n)}{\gcd( 2^r(q-1),n,u)}\right)=r+1-\nu_2(u)\le \nu_2((p-1)(j-i))=1+\nu_2(j-i),$$
and therefore $2^{r-\nu_2(u)}|(j-i)$. 

In other words, if the  coefficient of the monomial of degree $n-t-it$ is zero, then all the coefficients of the monomials of degree $n-t-jt$ with $j\equiv i\pmod {2^{r-\nu_2(u)}} $ are zero.
Thus, if $\lambda\notin \Lambda_u$, then  any coefficient of the form $x^{tj}$ is zero  and the weight  of $g_{\lambda}$ is $\frac nt$. Otherwise, 
 exactly $\frac nt\cdot \frac1{2^{r-\nu_2(u)}}$  coefficients of the monomials of the form $x^{tj}$ are zero, then  the weight  of $g_{\lambda}$ is $\frac nt\left(1-\frac 1{2^{r-\nu_2(u)}}\right)$, as we want to prove. \qed

\begin{corollary}\label{pares} Let $g$ be a polynomial  in the same condition of lemma \ref{weight}. Then
$$ \# \left\{ (\mu,\lambda) \in \F_q^2 \left| \omega (\mu g(x) + \lambda x^{t}g(x)) = \dfrac nt\left(1- \frac 1{2^{r-\nu_2(u)}}\right)\right.\right\} = 2^{r-\nu_2(u)}(q-1).$$
\end{corollary}
\proof
If $\mu = 0$ and $\lambda \ne 0$, then  $\omega (\lambda x^{t}g(x)) = \frac nt\left(1-\frac 1{2^{r-\nu_2(u)}}\right)$ and  we have $(q-1)$ ways to choose $\lambda$. 

Suppose that $\mu\ne 0$, then 
 $\omega(\mu g(x) + \lambda x^{t}g(x)) = \omega\left(g(x) + \frac{\lambda}{\mu}x^{t}g(x) \right),$ 
i.e. the weight only depends on the quotient $ \frac{\lambda}{\mu}$.
By  Lemma \ref{weight}  there exist  $2^{r-\nu_2(u)}-1$ values of  $ \frac{\lambda}{\mu}$ such that  $g(x) + \frac \lambda\mu x^tg(x)$ has weight $\frac nt\left(1-\frac 1{2^{r-\nu_2(u)}}\right)$, so we have $(q-1)(2^{r-\nu_2(u)}-1)$ pairs of this type.  
\qed

\begin{theorem} 
If $8|n$ and $q\equiv 3 \pmod 4$, then every irreducible code of lenght $n$ over $\F_q$  is one of the following class:
\begin{enumerate}[(a)]
\item A $[q;n, t, \frac nt]$-code, where  $4\nmid t$, $t|m'$  and  its weight enumerator is
$$A(z)=\sum_{j=0}^{t}  {t \choose j}(q-1)^j z^{j \frac nt}=(1+(q-1)z^{\frac nt})^t.
$$
\item A $[q;n,2t, d ]$-code, where  $t|m'$, $d=\frac nt(1-\frac 1{2^{r-\nu_2(u)}})$,  $0\le u\le r-2$ and  its weight enumerator  is 
$$A(z)=\left(1+2^{r-\nu_2(u)}(q-1)z^d+(q-1)(q+1-2^{r-\nu_2(u)})z^{\frac nt}\right)^t.$$
In particular,  if $\frac n{t2^{r-\nu_2(u)}}\nmid k$, then  $A_k=0$.
\end{enumerate}
\end{theorem}

\proof Observe that every irreducible code is generated by a polynomial of the form $\frac{x^n-1}{x^t-a}$ where $a\in \F_q$, or a polynomial of the form $g(x)=\frac{x^n-1}{(x^t-a)(x^t-a^q)}$, where $a$ satisfies the condition of lemma \ref{weight}.  In the first case, the result is the same of the Theorem \ref{4k+1}.  In  the second case, each codeword is of the form
$$\sum_{j=0}^ {2t-1} \lambda_j x^j g(x)=\sum_{j=0}^{t-1} h_j,$$
where $h_j=\lambda_j x^jg(x)+\lambda_{t+j} x^{t+j}g(x)$.
Since, for $0\le i<j\le t-1$, the polynomial $h_i$ and $h_j$  do not have non-null monomials of the same degree, it follows that
$$\omega\left(\sum_{j=0}^{t-1} h_j\right)=\sum_{j=0}^{t-1} \omega(h_j).$$

By Lemma  \ref{weight},  $h_j$ has weight  $\frac nt$, $d$ 
or $0$, for all  $j = 0, \dots, t-1$.
For each $j=0,1,\dots,t-1$,  
there exist   $(q^2-1)$ non-null pairs  $(\lambda_j, \lambda_{j+t})$ , and  by Corollary \ref{pares}, we know  that there exist  
 $2^{r-\nu_2(u)}(q-1)$ pairs with  weight $d$. 
Therefore, there exist 
$$q^2-1 -2^{r-\nu_2(u)} (q-1) = (q-1)(q+1-2^{r-\nu_2(u)})$$ pairs with weight $\frac nt$. 

So, in order to calculate  $A_k$, we need to select which  $h_l$'s have weight $d=\frac nt(1-\frac 1{2^{r-\nu_2(u)}})$ and which ones have weight $\frac nt$, so that the total weight is $k$.

If we chose $i$ of the first type and $j$ of the second type,  the first $h_l$'s can be
chosen by $ {t \choose i}\left(2^{r-\nu_2(u)}(q-1)\right)^i$ ways and for the other $t-i$ ones, there are 
$ {{t-i} \choose j}\left((q-1)(q+1-2^{r-\nu_2{u}})\right)^j$ ways of choosing $j$ with weight $\frac nt$. The remaining  $h_j$'s have weight zero.  Therefore
$$
 A_k = \sum_{k = di + \frac ntj\atop {0\le i+ j\le t}}{t \choose i}\left(2^{r-\nu_2(u)}(q-1)\right)^i{{t-i} \choose j}\left((q-1)(q+1-2^{r-\nu_2(u)})\right)^j,\\
$$
and
\begin{align*}
A(z)&=\sum_{0\leq i+ j\leq t}{t \choose i,j}\left(2^{r-\nu_2(u)}(q-1)z^d\right)^i\left((q+1-2^{r-\nu_2(u)})(q-1)^{j}z^{\frac nt}\right)^j\\
&=\left(1+2^{r-\nu_2(u)}(q-1)z^d+(q-1)(q+1-2^{r-\nu_2(u)})z^{\frac nt}\right)^t.
\end{align*}
In particular, the minimal distance  is $d$ and every non-null weight is divisible by $\gcd(d,\frac nt)=\frac n{t2^{r-\nu_2(u)}}$.
\qed  

\begin{remark}
As a direct consequence of Lemma \ref{factorst},  for all  $t$ positive divisor of  $m'$, 
there exist $2^{r-1-\nu_2(u)}\frac {\varphi(t)}{t} \gcd(n,q-1)$ irreducible cyclic $[q;n,t,d]$-codes if  $t$ is odd, and $2^{r-1}\frac {\varphi(t)}{t} \gcd(n,q-1)$  irreducible cyclic $[q;n,2t,\frac nt(1-\frac 1{2^r})]$-codes if $t$ is even.
\end{remark}

\begin{example} Let $q=31$ and $n=288=2^5\times 3$. Then $m'=3$, $l'=10$, $r=4$. If $h(x)$ denotes a irreducible factor of $x^{288}-1$, then $h(x)$ is a binomial of degree $1$, $2$, $3$ or $6$, or a  trinomial of degree $2$ or $6$.  The irreducible  codes generates by $\frac {x^{n}-1}{h(x)}$ (and therefore parity check polynomial $h$),  and its weight enumerators are showed in the following tables
\begin{center}
{\sc Codes generated by binomials}  \nopagebreak \\
\begin{tabular}{c|c|c}\hline
$[q;n,t,\frac nt]$-Code&$h(x)$&Weight enumerator\\ \hline\hline
$[31;288,1,288]$&$\scriptsize\begin{array}{c} x + 1\\ x + 5\\ x + 6\\  x + 25\\ x + 26\\ x + 30\end{array}$&$1+30z^{288}$ \\ \hline
$[31;288,2,144]$&$\scriptsize\begin{array}{c} x^2 + 1\\ x^2 + 5\\ x^2 + 25\end{array}$&$(1+30z^{144})^2$
 \\ \hline
$[31;288,3,96]$&$\scriptsize\begin{array}{c}x^3 + 5\\ x^3 + 6\\  x^3 + 25\\ x^3 + 26\end{array}$&$(1+30z^{96})^3$ 
 \\ \hline
$[31;288,6,48]$&$\scriptsize\begin{array}{c} x^6 + 5\\ x^6 + 25 \end{array}$&$(1+30z^{48})^6$
\\ \hline
\end{tabular}
\end{center}

\ 

\begin{center}{\sc Codes generated by trinomials of the form $x^2+ax+b$}  \nopagebreak \\
\begin{tabular}{c|c|c|c}\hline
$[q;n,2t,d]$-Code&$\nu_2(u)$&$h(x)$&Weight enumerator\\ \hline\hline
$[31;288,2,216]$&$2$&$\scriptsize\begin{array}{c}  x^2 + 8x +1\\  x^2 +9 x + 25 \\  x^2 +14x + 5 \\     x^2 +17x + 5 \\    x^2 +22 x + 25 \\      x^2 +23 x + 1 \\  \end{array}$& $1+120z^{216}+840z^{288}       $\\ \hline 
$[31;288,2,252]$&$1$&$\scriptsize\begin{array}{c}x^2 + x + 5\\ x^2 + 5x + 1\\   x^2 + 6x + 25\\ x^2 + 8x + 25\\  x^2 +9 x + 5 \\  x^2 +14 x + 1 \\     x^2 +17 x + 1 \\   x^2 +22 x + 5 \\      x^2 +23 x +25 \\    x^2 +25 x + 25 \\      x^2 +26 x + 1 \\        x^2 +30 x + 5 \\     \end{array}$& $1+240z^{252}+           720z^{288}$ \\ \hline 
$[31;288,2,270]$&$0$&$\scriptsize\begin{array}{c} x^2 +2 x + 5 \\ x^2 +4 x + 1 \\  x^2 +4 x + 5 \\  x^2 +7 x + 5 \\  x^2 +7 x + 25 \\  
   x^2 +8 x + 5 \\  x^2 +9 x + 1 \\  x^2 +10 x + 1 \\  x^2 +11 x + 1 \\  x^2 +11 x + 25 \\  x^2 +12 x + 25 \\   x^2 +14 x + 25 \\    x^2 +17 x + 25 \\     x^2 +19 x + 25 \\      x^2 +20 x + 1 \\        x^2 +20 x + 25 \\      x^2 +21 x + 1 \\       x^2 +22 x + 1 \\      x^2 +23 x + 5 \\    x^2 +24 x + 5 \\       x^2 +24 x + 25 \\       x^2 +27 x + 1 \\   x^2 +27 x + 5 \\    x^2 +29 x + 5 \\    \end{array}$&  $1+480z^{270}+480z^{288}          $\\ \hline 
\end{tabular}
\end{center}

\ 

\begin{center}{\sc Codes generated by trinomials of the form $x^6+ax^3+b$}  \nopagebreak \\
\begin{tabular}{c|c|c|c}\hline
$[q;n,2t,d]$-Code&$\nu_2(u)$&$h(x)$&Weight enumerator\\ \hline\hline
$[31;288,6,72]$&$2$&$\scriptsize\begin{array}{c}  x^6 + 9x^3 + 25 \\   x^6 + 14x^3 + 5 \\     x^6 + 17x^3 + 5 \\      x^6 + 22x^3 + 25 \\                              \end{array}$& $(1+120z^{72}+840z^{96})^3$
\\ \hline 
$[31;288,6,84]$&$1$&$\scriptsize\begin{array}{c}   x^6 + x^3 + 5\\     x^6 +6 x^3 + 25\\      x^6 +8 x^3 + 25\\     x^6 +9 x^3 + 5\\                                  x^6 +22 x^3 + 5\\    x^6 +23 x^3 + 25\\     x^6 +25 x^3 + 25\\      x^6 +30 x^3 + 5\\                               \end{array}$&  
$(1+240z^{84}+720z^{96})^3$
\\ \hline
$[31;288,6,90]$&$0$&$\scriptsize\begin{array}{c}   x^6 + 2x^3 + 5\\     x^6 + 4x^3 + 5\\        x^6 + 7x^3 + 5\\      x^6 + 7x^3 + 25\\                            x^6 + 8x^3 + 5\\    x^6 + 11x^3 + 25\\      x^6 + 12x^3 + 25\\        x^6 + 14x^3 + 25\\       x^6 + 17x^3 + 25\\     x^6 + 19x^3 + 25\\                             x^6 + 20x^3 + 25\\      x^6 + 23x^3 + 5\\       x^6 + 24x^3 + 5\\        x^6 + 24x^3 + 25\\       x^6 + 27x^3 + 5\\      x^6 + 29x^3 + 5\\                         \end{array}$&  
$(1+480z^{90}+480z^{96})^3$
\\ \hline
\end{tabular}
\end{center}
\end{example}

\end{document}